\documentclass[pra,aps,showpacs,twocolumn,superscriptaddress,]{revtex4}
\date{ }
\pacs{03.75.Hh, 67.85.-d}
\usepackage{graphicx}

\begin{document}

\title{Density ripples in expanding low-dimensional gases as a probe of correlations}

\author{A. Imambekov}
\affiliation{Department of Physics, Yale University, New Haven,
Connecticut 06520, USA}
\affiliation{Department of Physics and
Astronomy, Rice University, Houston, Texas 77005, USA}
\author{I. E. Mazets}
\affiliation{Atominstitut, Fakult\"at f\"ur Physik, TU-Wien,
Stadionallee 2, 1020 Vienna, Austria} \affiliation{Ioffe
Physico-Technical Institute, 194021 St.Petersburg, Russia}
\author{D. S. Petrov }
\affiliation{Laboratoire Physique Th\'{e}orique et Mod\'eles Statistique, Universit\'e Paris Sud, CNRS, 91405 Orsay, France}
\affiliation{Russian
Research Center Kurchatov Institute, Kurchatov
Square, 123182 Moscow, Russia}
\author{V. Gritsev}
\affiliation{Physics Department, University of Fribourg, Chemin
du Musee 3, 1700 Fribourg, Switzerland}
\author{S. Manz}
\affiliation{Atominstitut, Fakult\"at f\"ur Physik, TU-Wien,
Stadionallee 2, 1020 Vienna, Austria}
\author{S.~Hofferberth}
\affiliation{Department of Physics, Harvard University,
Cambridge, Massachusetts 02138, USA}
\author{T. Schumm}
\affiliation{Atominstitut, Fakult\"at f\"ur Physik, TU-Wien,
Stadionallee 2, 1020 Vienna, Austria} \affiliation{Wolfgang Pauli
Institute, University of Vienna, 1090 Vienna, Austria}
\author{E. Demler}
\affiliation{Department of Physics, Harvard University,
Cambridge, Massachusetts 02138, USA}
\author{J. Schmiedmayer}
\affiliation{Atominstitut, Fakult\"at f\"ur Physik, TU-Wien,
Stadionallee 2, 1020 Vienna, Austria}

\date{\today}

\begin{abstract}

We  investigate theoretically the   evolution of the two-point
density correlation function of a low-dimensional ultracold Bose
gas after  release from a tight transverse confinement. In the
course of expansion thermal and quantum fluctuations present in
the trapped systems transform into density fluctuations. For the
case of free ballistic expansion relevant to current experiments,
we present simple analytical relations between the spectrum of
``density ripples'' and the correlation functions of the original
confined systems. We analyze several physical regimes, including
weakly and strongly interacting one-dimensional (1D)  Bose gases
and two-dimensional (2D) Bose gases below the
Berezinskii-Kosterlitz-Thouless (BKT) transition. For weakly
interacting 1D Bose gases, we obtain an explicit analytical
expression for the spectrum of density ripples which can be used
for thermometry. For 2D Bose gases below the BKT transition, we
show that for sufficiently long expansion times the spectrum of
the density ripples has a self-similar shape controlled only by
the  exponent of the first-order correlation function. This
exponent can be extracted by analyzing the evolution of the
spectrum of density ripples as a function of the expansion time.
\end{abstract}

\maketitle

\def\be{\begin{equation}}
\def\ee{\end{equation}}
\def\bea{\begin{eqnarray}}
\def\eea{\end{eqnarray}}
\def\bean{\begin{mathletters}\begin{eqnarray}}
\def\eean{\end{eqnarray}\end{mathletters}}

\section{Introduction}

\subsection{Quantum noise studies of ultracold atoms}

Quantum correlations can be used to identify and study interesting
quantum phases and regimes in ultracold atomic systems. Recent
experimental advances include detection of the Mott insulator
phase of bosonic~\cite{Greiner} and fermionic~ \cite{bloch2}
atoms in optical lattices, production of correlated atom pairs in
spontaneous four-wave mixing of two colliding Bose-Einstein
condensates~\cite{asp1}, studies of dephasing \cite{deph-s} and
interference distribution functions \cite{tnoise-s} in coherently
split one-dimensional (1D)  atomic quasicondensates (QC),
observation of the Berezinskii-Kosterlitz-Thouless (BKT)
transition~\cite{Berezinskii, KT} in two-dimensional (2D)
quasicondensates~\cite{Zoran_KT}, and Hanbury-Brown-Twiss
correlation measurements for nondegenerate (ND) metastable $^4$He
\cite{Aspect} and $^3$He atoms~\cite{asp2}, bosonic~\cite{bloch1}
and fermionic~\cite{bloch_noise_fermions} atoms in optical
lattices, and in atom lasers~\cite{esslinger}. In one-dimensional
atomic gases~\cite{gorlitz,
1dexp,KWW,Paredes,Trebbia,Klaasjan,Bouchoule_review}, in situ
measurements of correlations have been attained by means of
photoassociation spectroscopy \cite{paw} or by measuring the
three-body inelastic decay \cite{tbl}, using the proportionality
of the corresponding rates to the zero-distance two-particle and
three-particle correlation functions, respectively \cite{kag85}.


Recently it was demonstrated that one can detect single neutral
atoms in a tight trap or
guide~\cite{em08,detector1,detector2,x1,x2,x3}. However,  direct
(not inferred from any kind of atomic loss rate~\cite{paw,tbl})
observation of interatomic correlations at short distances in
trapped ultracold atomic gases is hindered in many cases by
either the finite spatial resolution of the optical detection
technique or the very low detection efficiency of the scanning
electron microscope~\cite{em08}.
 Therefore one needs to
release ultracold atoms from the trap, diluting the atomic cloud
in the course of expansion.

In this paper we address  the question of how the correlations in
the low-dimensional system evolve during the time-of-flight
expansion, and discuss how the density variations in the
time-of-flight images relate to the properties of the original
trapped quantum gas.  These ``density ripples'' in the expanding
gas reflect the original thermal or quantum phase fluctuations
existing in the cloud under confinement. Such phase fluctuations
are already present in three-dimensional (3D) Bose-condensed
clouds under an external confinement with large aspect
ratio~\cite{Petrov2001}. Their effect on density ripples of
expanding clouds has been observed~\cite{Dettmer, Hellweg01,
Kreutzmann03}, but quantitative analysis of such experiments was
complicated since one had to take into account  interactions in
the course of expansion. However, for sufficiently strong
transverse confinement reached in current experiments with
low-dimensional gases (chemical potential of the order of the
transverse confinement frequency), the gas  expands rapidly in the
transverse direction  so interactions during the expansion stage
can be safely neglected. Then one can develop a simple analytical
theory, which directly relates the spectrum of the density ripples
after the expansion to the correlation functions of the original
fluctuating condensates. Similar question has been considered for
3D clouds expanding in the gravitational field but only for
noninteracting atoms~\cite{ballistic_3D}. We also note the
density ripples we discuss are different from the density
modulations which appear due to interactions during expansion and
have been studied in Refs.~\cite{Alon1} and~\cite{Alon2}.

\subsection{Density ripples in expanding condensates: Preview}

We consider one- or two-dimensional atomic gases released from a
tight trap formed by a scalar potential as realized on atom chips
or in optical lattice experiments. We consider the situation when
free expansion takes place in all three dimensions. This should
be contrasted to the expansion of such a gas inside a waveguide
\cite{KWW,Santos,gang,buljan,buljan2,buljan3,delCampo,Rigol,RigolOlshanii,GangardtMinguzzi},
with the transverse confinement being permanently maintained. In
the latter case, the nonlinear atomic coupling constant \bea
g_{1D}= 2\hbar \omega _\perp a_s, \eea where $\omega _\perp$ is
the transverse trapping frequency and $a_s$ is the atomic $s$-wave
scattering length, remains the same.  While a bosonic gas
rarifies during such expansion, collisions remain important. For
example, in the 1D case dynamics asymptotically reaches the
limiting Tonks-Girardeau (TG)~\cite{Gir} regime of impenetrable
bosons. In our case, if the fundamental frequency of the
potential of the transverse confinement is much larger than the
initial chemical potential of the atoms, the expansion in the
transverse directions  is determined mainly by the kinetic energy
stored in the initial localized state of the transverse motion.
Interatomic collisions play almost no role in the expansion.
Moreover tight transverse confinement decouples the motion of
trapped atoms in the longitudinal and transverse directions. Thus
when analyzing density ripples we can reduce the problem to the
same number of dimensions as  the initial trap (see discussion
below in Sec.~\ref{freeexp}). For a 1D trap we consider a
one-dimensional spectrum of density ripples, and for atoms which
were originally confined in a pancake trap we analyze
two-dimensional density ripples.

Before we consider a general formalism, it is useful to present
the analysis for the simplest situation. Let us assume that the
initial state can be described using the mean-field Bogoliubov
approach~\cite{Bogoliubov,Pitaevskii,Pethick}. Let $\hat
\Psi_{\vec k}^\dagger$ be the creation operator of atoms at
momentum $\vec{k}$  right before the expansion. 
After free expansion during time $t,$ in the Heisenberg
representation we have $\hat \Psi_{\vec k}^\dagger(t) = \hat
\Psi_{\vec k}^\dagger \, \, e^{i\frac{\hbar^2 k^2 t }{2m}},$
where $m$ is the atomic mass. Then the density operator at time
$t$ is given by
\begin{eqnarray}
\rho(\vec{r},t) = \frac1L\sum_{\vec{k}_1, \vec{k}_2} \hat
\Psi_{\vec{k}_1}^\dagger  \hat\Psi_{\vec{k}_2} e^{-i
(\vec{k}_1-\vec{k}_2)\vec{r}}
e^{\frac{it\hbar^2}{2m}(k_1^2-k_2^2)},
\end{eqnarray}
and for the density correlation function we obtain
\begin{eqnarray}
&&\langle \rho(\vec{r}_1,t) \rho(\vec{r}_2,t) \rangle  =
\frac{1}{L^2}\sum_{\vec{k}_1, \vec{k}_2,\vec{k}_3, \vec{k}_4}
\langle \hat\Psi_{\vec{k}_1}^\dagger \hat \Psi_{\vec{k}_2} \hat
\Psi_{\vec{k}_3}^\dagger \hat \Psi_{\vec{k}_4} \rangle
\nonumber\\
&\times&
e^{-i (\vec{k}_1-\vec{k}_2)\vec{r}_1}  e^{-i (\vec{k}_3-\vec{k}_4)\vec{r}_2}
 e^{\frac{it\hbar^2}{2m}(k_1^2-k_2^2)}
e^{\frac{it\hbar^2}{2m}(k_3^2-k_4^2)}.
\label{four_psi_eq}
\end{eqnarray}
The expectation value $\langle \hat\Psi_{\vec{k}_1}^\dagger
\hat\Psi_{\vec{k}_2}  \hat\Psi_{\vec{k}_3}^\dagger
\hat\Psi_{\vec{k}_4} \rangle $ should be taken in the original
condensate before the expansion. Within the  mean-field Bogoliubov
theory only  a state with $k=0$ is macroscopically occupied. Thus
in Eq.~(\ref{four_psi_eq}) we take two operators to be
$\sqrt{N}=\sqrt{n_{1D}L}$, where $n_{1D}$ is the atomic density
before the expansion. Thus Eq.~(\ref{four_psi_eq}) can be written
as
\begin{eqnarray}
 \langle \rho(\vec{r}_1,t) \rho(\vec{r}_2,t) \rangle  =
n_{1D}^2+ \frac{n_{1D}}{L} \, \sum_{\vec{q}\neq 0}  e^{i \vec{q}
(\vec{r}_1-\vec{r}_2)} \, \nonumber\\  \left[1+2\langle
\hat\Psi_{q}^\dagger \hat\Psi_{q} \rangle + \left(\langle
\hat\Psi_{-q} \hat\Psi_{q} \rangle+\langle \hat\Psi_{-q}^\dagger
\hat\Psi_{q}^\dagger \rangle \right)\cos{\frac{\hbar^2 q^2
t}{m}}\right].\label{eq4}
\end{eqnarray}
The Bogoliubov theory predicts expectation values of $\langle
\hat\Psi_{-q} \hat\Psi_{q} \rangle,  \langle
\hat\Psi_{-q}^\dagger \hat\Psi_{q}^\dagger \rangle,$ and
$1+2\langle \hat\Psi_{q}^\dagger \hat\Psi_{q} \rangle$ as
 \bea
\langle \hat\Psi_{-q} \hat\Psi_{q} \rangle= \langle
\hat\Psi_{-q}^\dagger \hat\Psi_{q}^\dagger
\rangle=-\frac{\mu}{2E_q}\left[1+2n_B\left(\frac{E_q}{k_BT}\right)\right],
\\
1+2\langle \hat\Psi_{q}^\dagger \hat\Psi_{q} \rangle=
\frac{\epsilon_q+\mu}{E_q}\left[1+2n_B\left(\frac{E_q}{k_BT}\right)\right],
 \eea
where $\epsilon_q= \hbar^2 q^2 /(2m),$ $\mu = g_{1D}n_{1D}$ is the
chemical potential, $E_q=\sqrt{\epsilon_q (2\mu+\epsilon_q)}$ is
the Bogoliubov excitation spectrum, and $n_B$ is the Bose
occupation number.

From these equations we can easily find the mean-field spectrum of
density ripples $\left \langle
 |\rho^{MF}(q)|^2\right\rangle$ [see Eq.~(\ref{rhoqeq}) and the discussion nearby
for the precise mathematical definition of the spectrum] \bea
 \left \langle
 |\rho^{MF}(q)|^2\right\rangle=n_{1D}\left[1+2n_B\left(\frac{E_q}{k_BT}\right)\right]\nonumber
 \\
\times
\left[\frac{\epsilon_q}{E_q}+\frac{\mu}{E_q}\left(1-\cos{\frac{\hbar^2
q^2 t}{m}}\right)\right]. \label{mean_field_final}
 \eea
The general character of the spectrum is clear from
Eq.~(\ref{mean_field_final}). As a function of momentum it is not
monotonic. We find minima near $\hbar q^2 t/m= 2\pi n,$  and
maxima close to $\hbar q^2 t/m= \pi (2n-1),$ where $n$ is a
positive integer number. Note that while the positions of maxima
and minima are essentially universal, the amplitude of individual
maxima depends on both temperature and interaction strength.

The mean-field analysis leading to Eq.~(\ref{mean_field_final})
is conceptually simple but has limited applicability. It is
applicable for weakly interacting 3D Bose condensates if the
interactions during expansion are switched off using Feshbach
resonances~\cite{Feshbach}. In lower dimensions, thermal and
quantum phase fluctuations are expected to suppress true long
range order in 2D Bose condensates at finite
temperature~\cite{MWH} and in 1D Bose condensates even at zero
temperature~\cite{Coleman}. In this paper we show how the analysis
of the density ripples can be extended to more complicated but
experimentally relevant situations when the mean-field approach
breaks down. We will find a similar structure to
Eq.~(\ref{mean_field_final}): positions of maxima and minima of
the spectrum are given by the same approximate universal
conditions on the momenta. However explicit expressions for the
strength of individual maxima will be very different. They will
contain rich information about fluctuations of low-dimensional
condensates.

\subsection{Relation to other work}

Conceptually, the question we consider in this  paper is somewhat
similar to the interpretation of cosmological observations. In
the latter case, quantum fluctuations present in the early
universe after its expansion result in observable anisotropies of
the cosmic microwave background radiation~\cite{Novikov,Nobel},
and in the density ripples of matter which eventually evolve into
galaxies~\cite{Universe}. In our case, density ripples of the
expanding clouds contain important information about correlations
present in the trapped state. Analogies between properties of
condensates and cosmology have attracted significant attention
recently~\cite{Garay, Volovik,FedichevFisher, Barcelo,
Uhlmann1,Hawking1,Hawking2,Uhlmann2}.

In addition to the mentioned approaches, several other techniques
have been used to experimentally study correlations of
low-dimensional gases. Some of them rely on creation of two
copies of the same cloud~\cite{HellwegCacciapuoti, Gerbier,
Richard, Hugbart, Clade}, while others require analysis of noise
correlations~\cite{PRA_noise} or in situ density-fluctuation
statistics~\cite{Esteve}. Interference experiments between two
low-dimensional clouds~\cite{deph-s,tnoise-s,Zoran_KT} can also
be used to characterize two-point and multi-point correlation
functions~\cite{pnas,Gritsev,interference_PRA,intrev}. Analysis
of density ripples is a much simpler experiment and, as we
discuss in this paper, can be used for thermometry. This is
particularly important for weakly interacting 1D Bose
quasicondensates~\cite{Popov,Petrov2000}, for which  the standard
approach to measuring temperature by fitting density profiles
cannot be extended to temperatures of the order of the chemical
potential. In this regime, the chemical potential is very weakly
dependent on the temperature~\cite{MoraCastin}, thus finite
temperature leads only to small corrections to the ``inverted
parabola'' density profile~\cite{Pollet}. An improved thermometry
method based on comparison of  in situ measured density profiles
with solutions of Yang and Yang equations~\cite{YangYang} in the
local-density approximation has been developed in
Ref.~\cite{Klaasjan}.

There has been  significant theoretical interest in correlation
functions of the 1D Bose gas. At distances much larger than the
healing length, correlation functions are described by Luttinger
liquid theory \cite{EL,Haldane,Caza04}. In the weakly interacting
quasicondensate regime, correlation functions can be described by
extension of Bogoliubov theory to low-dimensional
gases~\cite{Petrov2000,Stoof_PRA, Stoof_PRL,
MoraCastin,Petrov_review}. In the strongly interacting regime,
one can use ``fermionization''~\cite{Gir} of a  1D Bose gas to
evaluate correlation functions at all distances as certain
determinants~\cite{g1TG}. The Lieb-Liniger model~\cite{LL} which
describes the 1D Bose gas is exactly solvable, and one can also
analytically obtain zero-distance two-point~\cite{g2first,KGDS03}
and three-point~\cite{g3} density correlations for any interaction
strength, and extract certain dynamical correlation
functions~\cite{IG_PRL08, CauxDSF,CauxG,GRD,Cherny} from the exact
solution. Various numerical techniques have been used as
well~\cite{Pollet,asg, Drummond,Dalfovo}, and recent results
including the decoherent quantum regime \cite{KGDS03,dqr} are
summarized in Refs.~\cite{g2last} and~\cite{Deuar}.

Two-dimensional systems have also been a subject of considerable
experimental
~\cite{gorlitz,Zoran_KT,Clade,Hadzi_NJP,2dconds,Kruger,JILA} and
theoretical work
\cite{Kadanoff,Popov,Petrov2D_PRL,Petrov_review,2DRMP,Holzmann,Prokofiev,Prokofiev2,fernandez,Gies,hutchinson}.

\subsection{Structure of the paper}

This paper is organized as follows. In Sec.~\ref{freeexp} we
derive simple analytical relations between the density ripples
after the expansion and the correlation functions of the original
system before the expansion.  In Sec.~\ref{weakinter} we analyze
the case of weakly interacting 1D Bose gases and obtain  explicit
expression for the spectrum of density ripples. In
Sec.~\ref{TGsection} we consider the case of a strongly
interacting 1D Bose gas. In Sec. \ref{1Dgas} we review general
features of the density-density correlation function in expanding
1D Bose clouds. In Sec.~\ref{2Dsection} we discuss 2D Bose systems
below the BKT transition~\cite{Berezinskii, KT}. We summarize our
results and make concluding remarks in Sec.~\ref{conclusions}.

\section{Free expansion}
\label{freeexp}

In this section we focus on the atoms expanding from a
one-dimensional trap. The atom field operator evolution during the
free expansion is given by \cite{FH}
\begin{equation}
\hat{\Psi }({\bf r},t)=\int d^3{\bf r}^\prime \, G_3({\bf r}-{\bf r}^\prime ,t)
\hat{\Psi }({\bf r}^\prime ,0),
\label{eq:1}
\end{equation}
where the Green's function of free motion is
\begin{equation}
G_3({\bf r}-{\bf r}^\prime ,t)=G_1(x-x^\prime ,t) G_1(y-y^\prime ,t)
G_1(z-z^\prime ,t) ,
\label{eq:2}
\end{equation}
\begin{equation}
G_1(\xi ,t)= \sqrt{\frac m{2\pi \, i \, \hbar t}}\exp \left(
i\frac {m\xi ^2}{2\hbar t} \right) ,
\label{eq:3}
\end{equation}
with $m$ being the atomic mass. Tight transverse confinement
decouples the motion of trapped atoms in the $(y,z)$-plane and
along the waveguide axis $x$, so that the transverse motion is
confined to its ground state $f_\perp (y,z),$ and $\hat{\Psi }
({\bf r},0) =f_\perp (y,z)\hat{\psi }(x,0)$. This, alongside with
Eq. (\ref{eq:2}), allows for a separation of motion in the
longitudinal and transverse directions, effectively reducing the
problem to 1D.

We introduce the  two-particle density matrix for the longitudinal
motion as
\begin{equation}
\rho (x_1,x_2;x_1^\prime ,x_2^\prime ;t) =\left \langle \hat{\psi }^\dag
(x_1 ^\prime ,t) \hat{\psi }^\dag (x_2 ^\prime ,t) \hat{\psi }(x_2 ,t)
\hat{\psi }(x_1 ,t)  \right \rangle .
\label{eq:4}
\end{equation}
Then  we define the two-point density correlation function  \bea
g_2(x_1,x_2 ;t)= \frac{\rho (x_1,x_2;x_1,x_2
;t)}{n(x_1,t)n(x_2,t)}, \label{g2def} \eea
  where $n(x,t) = \left
\langle \hat{\psi }^\dag (x,t)\hat{\psi }(x,t) \right \rangle $.
The free evolution of the two-particle density matrix is  given
by the convolution of the two-particle density matrix at $t=0$
with four respective Green's functions, one for each spatial
argument, two of the Green's functions being complex conjugate.
We are interested in the case $x_1=x_1^\prime $ and
$x_2=x_2^\prime $ in the final state. Then we obtain
\begin{eqnarray}
&&\rho (x_1,x_2;x_1,x_2 ;t)=\nonumber \\ &&\int dx_3 \int dx_3^\prime
\int dx_4 \int dx_4^\prime G_1(x_1-x_3,t) G_1(x_2-x_4,t) \nonumber \\
&& \times G_1^*(x_1-x_3^\prime ,t) G_1^*(x_2-x_4^\prime ,t)
\rho (x_3,x_4;x_3^\prime ,x_4^\prime  ;0).
\label{eq:5}
\end{eqnarray}

We assume that the product of the typical velocity of the atoms
in the $x$-direction and the expansion time is much smaller than
the size of the trapped atomic cloud. Then we are allowed to
consider a uniform sample with  length $L\rightarrow \infty $,
with the 1D number density $n_{1D}=N/L$ being kept constant in
the thermodynamic limit ($N$ being the total number of atoms).
 Note that this limit is
opposite to  the conventional limit of infinitely large expansion
times, in which density in real space reflects the initial
momentum distribution (in that regime, it was recently
proposed~\cite{MAV} that noise correlations in density profiles
can be used to probe properties of low-dimensional gases).

 In our limit $n(x,t)= n_{1D}$ is constant in time, and the
two-particle density matrix is translationally invariant (it does
not change if all four of its spatial arguments are shifted by the
same amount) at any time. The density  correlation function then
depends on the coordinate difference only, so we use the notation
 \bea g_2(x_1-x_2;t)\equiv g_2(x_1,x_2;t).\eea
 Using the translational invariance of the two-particle density matrix and
the identity
 \bea \delta(x)=\frac1{2\pi}\int_{-\infty }^\infty dy \exp\left(i y x \right),\eea
 we arrive  at
\begin{eqnarray}
&&\varrho (x_1-x_2;x_1-x_2 ;t)=\frac m{4\pi \hbar t} \nonumber \\
&& ~\times \int _{-\infty }^\infty dx \int _{-\infty }^\infty dx^\prime
\exp \left \{ i\frac m{4\hbar t} \left[ (x_1-x_2-x)^2 \right. \right. \nonumber \\
&& ~\left. \left. - (x_1-x_2-x^\prime )^2 \right] \right \}
\varrho \left(  x;{x^\prime };0\right) ,\label{eq:6}
\end{eqnarray}
where
\begin{equation}
\varrho (x;x^\prime ;t)\equiv
\rho \left( \frac x2,-\frac x2;\frac {x^\prime }2,-\frac {x^\prime }2;t\right) .
\label{eq:7}
\end{equation}
Obviously, $\rho (x_1,x_2;x_1,x_2 ;t)=\varrho
(x_1-x_2;x_1-x_2;t)=n_{1D}^2g_2(x_1-x_2;t).$ The physical meaning
of Eq. (\ref{eq:6}) is that the motion of the center of mass of an
atomic pair plays no role in the dynamics of establishing
$g_2(x_1-x_2;t)$, which is fully determined by the relative
motion. The relative-motion degree of freedom is characterized by
the reduced mass $m/2$~\cite{mnote}.

Let us now consider some properties of the two-particle density
matrix  $\varrho (x_1;x_2;t)$ for bosons. Changing the sign of
$x_1$ or $x_2$ is equivalent to a permutation of two bosons and,
hence, does not change the two-particle  density matrix, i.e.,
\begin{equation}
\varrho (x_1;x_2 ;t)=\varrho (|x_1|;|x_2| ;t) .
\label{eq:8}
\end{equation}
For the regimes we consider the density matrix of neutral bosons
can be assumed to be real. This, together with the Hermicity
property, results in
\begin{equation}
\varrho (x_1;x_2;t)=\varrho (x_2;x_1;t) .
\label{eq:9}
\end{equation}
Using Eqs. (\ref{eq:8}) and  (\ref{eq:9}) and Fourier transforming
the Green's functions, we can reduce Eq. (\ref{eq:6}) to
\begin{eqnarray}
\varrho (x;x;t)&=&\frac 2\pi \int _0^\infty dq\int _0^\infty dX\cos qx\cos qX
\nonumber \\
&& \times \varrho (|X-\frac{\hbar qt}{m}|;|X+\frac{\hbar
qt}{m}|;0). \label{eq:10}
\end{eqnarray}
Alternatively, this equation can be written as
 \bea &\left \langle
|\rho(q)|^2\right \rangle =\int _{-\infty}^{\infty} dX \cos qX \nonumber\\
&\times \left \langle \hat{\psi }^\dag (\frac{\hbar qt}{m},0)
\hat{\psi }^\dag (X ,0)  \hat{\psi }(X+\frac{\hbar qt}{m} ,0)
\hat{\psi }(0,0)
 \right \rangle.
 \label{rhoqeq}
 \eea
Here $\langle |\rho(q)|^2 \rangle$ is the spectrum of density
ripples at time $t,$ which in experiment can be obtained by
Fourier transforming absorption images  after
expansion~\cite{positivitynote}. It is related to two-point
density correlation function as~\cite{minus1note}
 \bea
\left \langle |\rho(q)|^2\right \rangle = n_{1D}^2
\int_{-\infty}^{\infty} dx \exp{\left(i q
x\right)}\left(g_2(x;t)-1\right). \label{rhovsg2}
 \eea
Equations (\ref{eq:10}) and (\ref{rhoqeq})
provide a simple analytical relation between
the properties of the density ripples after the expansion and the
correlation functions before the expansion.

It is straightforward to generalize the above analysis  to the 2D
case. In particular, the analog of Eq.~(\ref{rhoqeq}) for the
time evolution of the two-point density correlation function has
the same form, with $X$ substituted by ${\bf r}$ and $q$ treated
as a 2D vector. Namely, we obtain
 \bea
&\left \langle
|\rho({\bf q})|^2\right \rangle =\int_{R^2} d^2{\bf r} \cos{\bf q r}  \nonumber\\
&\times \left \langle \hat{\psi }^\dag (\frac{\hbar {\bf
q}t}{m},0) \hat{\psi }^\dag ({\bf r} ,0)  \hat{\psi }({\bf r
}+\frac{\hbar {\bf q}t}{m} ,0) \hat{\psi }(0,0) \right \rangle.
\label{rhoqeq2}
 \eea


%
%
%

\section{1D Bose gases}

\subsection{Weakly interacting 1D Bose gases}
\label{weakinter}

In this subsection we will consider the spectrum of density
ripples of weakly interacting 1D quasicondensates.
Before we proceed to full analytical theory, let us return to the
simple  mean-field Bogoliubov approach, which we discussed in the
introduction. Readers may be skeptical about the applicability of
the mean-field approach to 1D. Indeed,  there is no long-range
order for 1D gases even at zero temperature~\cite{Coleman}, and
mean-field approach is generally not applicable. However, in the
regime of weak interactions and under certain conditions on the
expansion time $t,$ the  spectrum of density ripples is captured
correctly by the mean-field approach, as we will verify later in
a rigorous calculation.

We consider Eq.~(\ref{mean_field_final}) in the limit
$\epsilon_q\ll~\mu,  E_q\ll k_BT.$  In this case one can neglect
the first term in the second parentheses, use an approximation
$E_q\propto q,$ and expand the Bose occupation number, leading to
\bea
  \frac{ \left\langle
 |\rho^{MF}(q)|^2\right\rangle}{n_{1D}^2}\approx \frac{2mk_BT \left(1-\cos{\frac{\hbar^2
q^2 t}{m}}\right)}{\hbar^2 n_{1D} q^2}.
\label{mean_field_d1}
\eea

Let us now present a full calculation, which does not make a
mean-field approximation. For weak interactions Bogoliubov theory
has been extended to low-dimensional
quasicondensates~\cite{MoraCastin}, and can be used to calculate
correlation functions at all distances. For quasicondensates, the
fluctuations of the phase are described by the Gaussian action.
For Gaussian actions, higher order correlation functions are
simply related to two-point correlation functions (see, e.g.,
Refs.~\cite{GNT} and~\cite{Giamarchibook}), and  the four-point
correlation function in Eq.~(\ref{rhoqeq}) factorizes into
products of two-point correlation functions of bosonic fields
as~\cite{intrev}
 \bea \left \langle \hat{\psi }^\dag (x_1 ^\prime ,0)
\hat{\psi }^\dag (x_2 ^\prime ,0) \hat{\psi }(x_1 ,0)
\hat{\psi}(x_2 ,0)  \right \rangle = \nonumber \\
\frac{\prod_{i,j=1}^{2}\left \langle  \hat{\psi }^\dag (x_i
^\prime ,0) \hat{\psi }(x_j ,0)\right \rangle}{\left \langle
\hat{\psi }^\dag (x_1 ^\prime ,0) \hat{\psi }(x_2^\prime
,0)\right \rangle \left \langle  \hat{\psi }^\dag (x_1,0)
\hat{\psi }(x_2,0)\right \rangle}. \label{factor}
 \eea
 This equation gives correct result for all values of $x_1,x_2,x_1',$ and $x_2'$
 in the leading order over $1/K$ expansion (see the definition of $K\gg 1$
 below), and can be used to evaluate the spectrum of the density ripples
 in weakly interacting condensates for all times. The two-point correlation function $\left \langle  \hat{\psi
}^\dag (x_1 ^\prime,0) \hat{\psi }(x_1 ,0)\right \rangle = n_{1D}
g_1(x_1^\prime-x_1) $ is translationally invariant, and is simply
related to predictions of the Bogoliubov theory.  For 1D
quasicondensates, one has~\cite{MoraCastin}
 \bea g_1(x)=\frac{\left \langle  \hat{\psi
}^\dag (x,0) \hat{\psi }(0,0)\right
\rangle}{n_{1D}}=\exp{\left[-\frac1{2K}f\left(\frac{x}{\xi_h}\right)\right]},
 \eea
where $K=\pi \hbar n_{1D}/(mc) \gg 1$ is the Luttinger liquid
parameter, $c$ is the speed of sound, $\xi_h=\hbar/\sqrt{m \mu}$
is the healing length, and $\mu$ is the chemical potential. When
only lowest transverse mode is occupied ($\mu\ll \hbar
\omega_\perp$), speed of sound is given by $c=\sqrt{2\hbar
\omega_\perp n_{1D} a_s/m}.$ The dimensionless function $f(s)$
depends on the temperature, and equals
 \bea
 f(s)=2\int_{0}^{\infty}dk(1-\cos{k s}) \left\{\right[u_k^2+v_k^2\left] n_k + v_k^2
 \right\},
 \eea
where
 \bea
u_k=\frac12\left[ \left(\frac{k^2 + 4
}{k^2}\right)^{1/4}+\left(\frac{k^2}{k^2 + 4
}\right)^{1/4}\right],\label{17}\\ v_k=\frac12\left[
\left(\frac{k^2}{k^2 + 4
}\right)^{1/4}-\left(\frac{k^2 + 4 }{k^2}\right)^{1/4}\right],\\
n_k=\frac1{\exp{\left(\frac{\mu\sqrt{k^2(k^2+4)}}{2 k_B
T}\right)}-1}. \label{19} \eea For finite temperatures, the
function $f(s)$ has the following asymptotic behavior \bea
f(s)\approx \pi |s| \frac{k_B T}{\mu} + C \;\; \mbox{for} \;\;
\pi |s|\frac{k_B T}{\mu}\gg 1, \label{assymp} \eea where $C\equiv
C(k_B T/\mu)$ is of order $O(1)$ for $k_B T\sim \mu.$

Quasicondensate theory is valid \cite{g2last, Deuar, MoraCastin}
for temperatures
 \bea k_B T/\mu \ll K/\pi, \label{acond}\eea
significantly beyond the regime of validity of Luttinger liquid
theory, which is restricted to $k_B T/\mu\ll 1.$ The longitudinal
density profile of a quasicondensate in external harmonic
confinement follows the inverted parabola shape under condition
(\ref{acond}), see, e.g., Ref.~\cite{Pollet}. Due to the low
fraction of the thermally populated excited states, it is
problematic to extract the temperature of the gas from fitting
bimodal distributions to the observed density profiles.
Below we show that the spectrum of density ripples can be used as
a convenient tool to characterize the temperature, and is
sensitive to temperatures of the order of the chemical potential.

\begin{figure}

\vspace*{3mm}
\includegraphics[width=8.5cm]{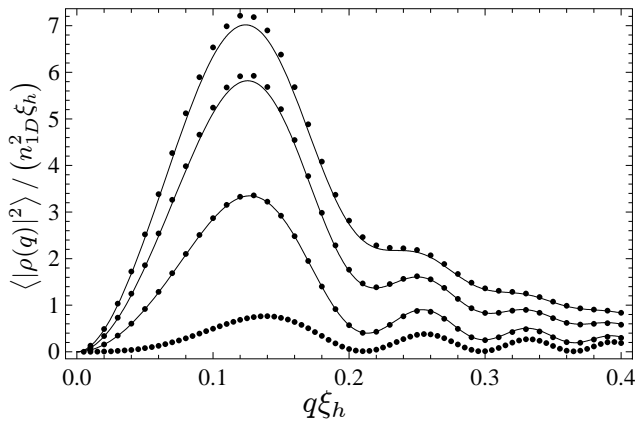}
\vspace*{3mm}

\caption{\label{TP} Normalized spectrum of density ripples $\left
\langle|\rho(q)|^2\right \rangle/(n_{1D}^2 \xi_h)$ for weakly
interacting 1D quasicondensate of $^{87}$Rb atoms with density
$n_{1D}=40\,\mbox{$\mu$}\mbox{m}^{-1},$ transverse confinement
frequency $\omega_\perp=2\pi \times 2\,\mbox{kHz},$ Luttinger
liquid parameter $K\approx 47,$ and healing length $\xi_h\approx
0.37\,\mbox{$\mu$}\mbox{m}.$ Expansion time is fixed at
$t=27\,\mbox{ms}$ (with $\frac{1}{\xi_h}\sqrt{\frac{\hbar
t}{m}}\approx 11.8$), and temperatures equal (top to bottom)
$T=40\,\mbox{nK}\, (k_BT/\mu=1),\; T=27\,\mbox{nK}\,
(k_BT/\mu=0.67),$ $T=12\,\mbox{nK}\, (k_BT/\mu=0.3), $ and $T=0.$
Values on the axes of this and subsequent plots are dimensionless.
Dots are obtained by numerical integration of Eq.~(\ref{rhoqeq})
in the weakly interacting limit making use of
Eqs.~(\ref{factor})-(\ref{19}).
 Solid lines
correspond to analytical results [Eq.~(\ref{rhoqfinal})], which
are derived under condition (\ref{analcond}).}
\end{figure}

\begin{figure}
\vspace*{3mm}
\includegraphics[width=8.5cm]{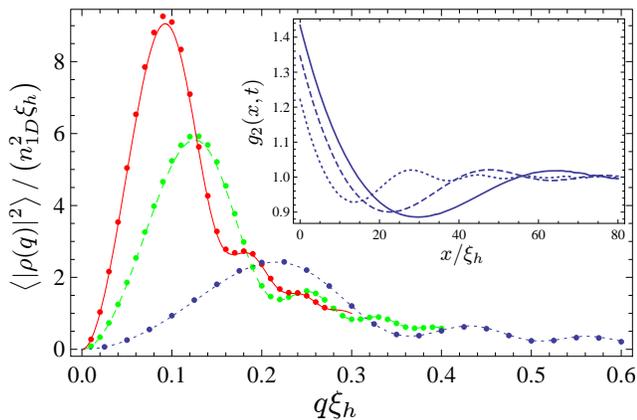}
\vspace*{3mm}

\caption{\label{TF}(Color online) Normalized spectrum of density
ripples $\left \langle|\rho(q)|^2\right \rangle/(n_{1D}^2 \xi_h)$
with the same parameters as in Fig.~\ref{TP} but for a fixed
temperature  $T=27\mbox{nK}\, (k_BT/\mu=0.67),$ and various times
of flight: $t=49\mbox{ms}$ (red, solid), $t=27\mbox{ms}$ (green,
dashed), and $t=9.5\mbox{ms}$ (blue, dotted). Dots are obtained
by numerical integration of Eq.~(\ref{rhoqeq}) in the weakly
interacting limit making use of Eqs.~(\ref{factor})-(\ref{19}).
Lines correspond to analytical results [Eq.~(\ref{rhoqfinal})],
which are derived under condition (\ref{analcond}). Inset shows
$g_2(x;t),$ obtained from $\left \langle|\rho(q)|^2\right \rangle$
using the inverse of Eq.~(\ref{rhovsg2}). }
\end{figure}

To be specific, let us consider the case of $^{87}$Rb atoms
(scattering length $a_s=5.2\,\mbox{nm}$) with density
$n_{1D}=40\,\mbox{$\mu$}\mbox{m}^{-1}$ and transverse confinement
frequency $\omega_\perp=2\pi\times2\,\mbox{kHz},$ resulting in
Luttinger liquid parameter $K\approx 47$ and healing length
$\xi_h\approx 0.37\,\mbox{$\mu$}\mbox{m}.$ We can use
Eqs.~(\ref{factor})-(\ref{19}) to numerically evaluate in-trap
correlation functions. By performing then a numerical integration
of Eq.~(\ref{rhoqeq}) for various temperatures and expansion
times, we can evaluate the spectrum of density ripples under
condition (\ref{acond}), and the results are shown in Figs.
\ref{TP} and \ref{TF}. In the inset to Fig.~\ref{TF} we also show
$g_2(x;t)$ evaluated using the inverse of Eq.~(\ref{rhovsg2}). In
the quasicondensate regime the behavior of $g_2(x;t)$ follows the
qualitative discussion  of Sec.~\ref{1Dgas}.

There are several qualitative features that should be noted. The
spectrum of density  ripples is not a monotonic function, and can
also have several maxima.  The positions of the maxima only
weakly depend on the temperature, and are mostly determined by the
expansion time. The amplitude of the ripples, on the other hand,
significantly depends both on the expansion time and temperature.

Let us now derive a simple analytical expression for the spectrum
of density ripples, which is valid in the regime [justified below
after Eq.~(\ref{extrrule})]
 \bea
 \frac{\pi}{\xi_h}\sqrt{\frac{\hbar t}{m}}\frac{k_BT}{\mu} \gg 1
 \label{analcond}.
 \eea
Under this condition one can use Eq.~(\ref{assymp}) and
approximate the two-point correlation function by
 \bea
 g_1(x)\approx \exp{\left(-|x|/\lambda_T\right)} \; \mbox{for} \; |x|\gg \xi_h\frac{\mu}{k_B T}, \label{g1form}
 \eea
where $\lambda_T$ is defined by
 \bea
 \lambda_T=\frac{2K \xi_h \mu}{\pi k_B T}=\frac{2\hbar^2 n_{1D}}{m k_B T}, \label{ltdef}
 \eea
and does not depend on interaction strength, as long as
Eq.~(\ref{acond}) is satisfied.

Using Eqs.~(\ref{factor}) and (\ref{g1form}), the second line of
Eq.~(\ref{rhoqeq}) can be written as
 \bea
 \frac{g_1(\hbar q t/m)^2
g_1(X)^2}{g_1(X-\hbar q t/m)g_1(X+\hbar q t/m)} \approx
\exp{\frac{-X}{\lambda_T}}  \; \mbox{for}\; X \leq \frac{\hbar q t}{m},\nonumber \\
\mbox{and}\; \exp{\frac{-\hbar q t}{m\lambda_T}}
\;\mbox{otherwise}.\nonumber
 \eea
The constant term $\exp{\frac{-\hbar q t}{m\lambda_T}}$ is
responsible for $g_2(x\rightarrow \infty,t)=1.$ Since according to
Eq.~(\ref{rhoqeq}) we need to take a Fourier transform of the
above expression, subtracting the constant on the whole interval
$(0,\infty)$ does not affect $\left \langle |\rho(q)|^2\right
\rangle$ for $q\neq 0,$ and  we obtain
 \be
\frac{ \left \langle |\rho(q)|^2\right \rangle}{n_{1D}^2} \approx
2 \int_0^{\frac{\hbar q t}{m}} dx \cos{q x}
\left(\exp{\frac{-x}{\lambda_T}}-\exp{\frac{-\hbar q
t}{m\lambda_T}}\right).\label{int} \ee This integral can be
evaluated in a closed form, and leads to an analytical answer
 \be
\frac{ \left \langle |\rho(q)|^2\right
\rangle}{n_{1D}^2\xi_h}\approx\frac{\lambda_T q -e^{\frac{-2\hbar q t
}{m\lambda_T}}\left(\lambda_T q \cos{\frac{\hbar q^2
t}{m}}+2\sin{\frac{\hbar q^2 t}{m}}\right)}{q
\xi_h\left(1+\lambda^2_T q^2\right)}. \label{rhoqfinal}
 \ee
 Note that the last equation reduces to Eq.~(\ref{mean_field_d1}) when $\lambda_T q \gg 1$ and $
 \hbar q t/m\lambda_T \ll 1$.
 Figures~\ref{TP} and~\ref{TF} show an excellent agreement between
the analytical result and numerical integration described earlier
after Eq.~(\ref{acond}). The analytical result shows the same
non-monotonic behavior as the numerical calculations. The
parameter $\lambda_T$  defines a time scale
 \bea
 t_c\approx 6.5 \frac{m \lambda_T^2}{\hbar}
 \eea
 after which only a single maximum persists.
When several maxima and minima are present, their positions can
be estimated by
 \bea \frac{\hbar q^2 t}{m}\approx
\pi(2n-1/2\mp1/2),\label{extrrule}
 \eea
where the upper (lower) sign corresponds to the $n$th maximum
(minimum). These conditions can be understood as a ``standing
wave'' conditions in Eq.~(\ref{int}), and become more precise at
lower temperatures.

 The appearance of minima and maxima in the
spectrum of density ripples can be understood in terms of
matter-wave near-field diffraction. The analogous effect for
light waves (in the spatial domain) is known as the Talbot
effect~\cite{Talbot}. Its matter-wave counterpart has been also
observed in diffraction of atoms on a grating~\cite{Talbotnew}.
In our case, we observe near-field diffraction in the time
domain. For each expansion time, a certain momentum contribution
will be ``imaged'' onto itself, leading to a minimum in the
spectrum of density ripples for a given momentum $q.$ As compared
to diffraction on a regular grating with a fixed period, the
typical fluctuation length in the trapped cloud is not constant,
but distributed around the thermal length $\lambda_T$. Therefore,
minima in the spectrum appear for any sufficiently small expansion
time, according to condition (\ref{extrrule}).

Condition (\ref{analcond}) can now  be  justified in the regime
where $\left \langle |\rho(q)|^2\right \rangle$ is near its
largest values. In such case most of the contributions to
Eq.~(\ref{rhoqeq}) come from distances of the order $\sqrt{\hbar
t/m},$ and Eq.~(\ref{analcond}) follows from Eq.~(\ref{acond}).

So far we have been assuming that the quasicondensate is deep in
the 1D regime, $\mu, k_BT \ll \hbar \omega_\perp.$ While
Eqs.~(\ref{factor})-(\ref{19}) are valid only under such
assumption, Eqs.~(\ref{g1form}) and (\ref{ltdef}) also work in
the weakly interacting quasi-1D regime,
 \bea
 \mu, k_BT \sim \hbar \omega_\perp.
 \eea
Indeed, they rely only on the 1D nature of long-range
correlations, weakness of interactions, and the property $c K =\pi
n_{1D}/m,$ which is a consequence of the Galilean
invariance~\cite{Haldane}.  In Eqs.~(\ref{acond}) and
(\ref{analcond}), the Luttinger liquid parameter $K$ can be
obtained as $K=\hbar \pi n_{1D}/(m c),$ where the square of the
sound velocity $c$ can be determined from compressibility as
$c^2=n_{1D}(\partial{\mu}/\partial{n_{1D}})/m.$ For chemical
potential $\mu,$ one can use an approximate
relation~\cite{Gerbier_t} $\mu=\hbar \omega_\perp(\sqrt{1+4 a_s
n_{1D}}-1).$

Let us now briefly review the conditions under which one can
neglect interactions in expanding 1D clouds  and the effects of
finite condensate length $L.$ Transverse expansion takes place at
the times of the order of inverse transverse confinement
$\omega_{\perp}^{-1}.$ Up to the times of this order, one cannot
neglect interactions during the expansion. Correlation functions
which enter Eq.~(\ref{rhoqeq}) will be smeared up to the
distances of the order $\delta x\sim c/\omega_{\perp}=\xi_h
\mu/\omega_{\perp},$ and smearing will only weakly affect the
final result for $\left \langle |\rho(q)|^2\right \rangle$ if
$q\delta x \ll 1.$ Thus to observe an oscillating spectrum of
density ripples, one needs to satisfy the  condition
 \bea
\xi_h\sqrt{\frac{m}{\hbar t}} \frac{\mu}{\hbar \omega_{\perp}}\ll
1,
 \eea
which easily holds for the parameters shown in Figs.~\ref{TP}
and~\ref{TF}. In addition, one can use Feshbach
resonances~\cite{Feshbach} to completely switch off interactions
during the expansion.

Locally, corrections due to finite $L$ can be neglected, if finite
limits of integration in Eq.~(\ref{eq:6}) lead to smearing of
delta  functions up to the distances at which the correlation
functions change considerably. This change can occur either
because of the variations of the density in external confinement
at distances $\sim L,$ or because of the decay of correlations
for finite temperatures at distances of the order $\sim K
\xi_h/a.$ Thus for finite temperatures these conditions read as
 \bea \frac{m L}{\hbar t} \min{(L,K \xi_h/a)} \gg1, \label{finiteL} \eea
 and
are easily satisfied for parameters considered earlier, and, e.g.,
longitudinal frequency $\omega_x=2\pi \times 5\; \mbox{Hz}.$
Under condition (\ref{finiteL}) one can take the inhomogeneity of
the density profile into account within the local-density
approximation by averaging the prediction of
Eq.~(\ref{rhoqfinal}).

\subsection{Strongly interacting 1D Bose gases}
\label{TGsection}

Let us now describe the evolution of the two-point density
correlation function $g_2(x;t)$ of a strongly interacting 1D Bose
gas. A dimensionless parameter which controls the strength of
interactions at zero temperature can be written as
 \bea
 \gamma=\frac{mg_{1D}}{\hbar^2 n_{1D}}=\frac{2m\omega_\perp a_s}{\hbar
 n_{1D}} \gg 1.
 \eea
Under such conditions, the bosonic wave function takes on fermion
properties, and the density correlation function in the trap
$g_2(x;0)$ is the same as for non-interacting fermions of the
same density and temperature. In particular, it vanishes at
$x=0,$ and one has $g_2(0;0)=0.$ However, the correlation
functions that contain creation and annihilation operators at
different points, such as $\varrho(x_1;x_2;0)$ in
Eq.~(\ref{eq:10}), are not the same as for non-interacting
fermions. This happens because bosonic operators, when written in
terms of fermionic operators, contain a ``string'' which ensures
proper commutation relations.

\begin{figure}

\vspace*{3mm}

\includegraphics[width=8.5cm]{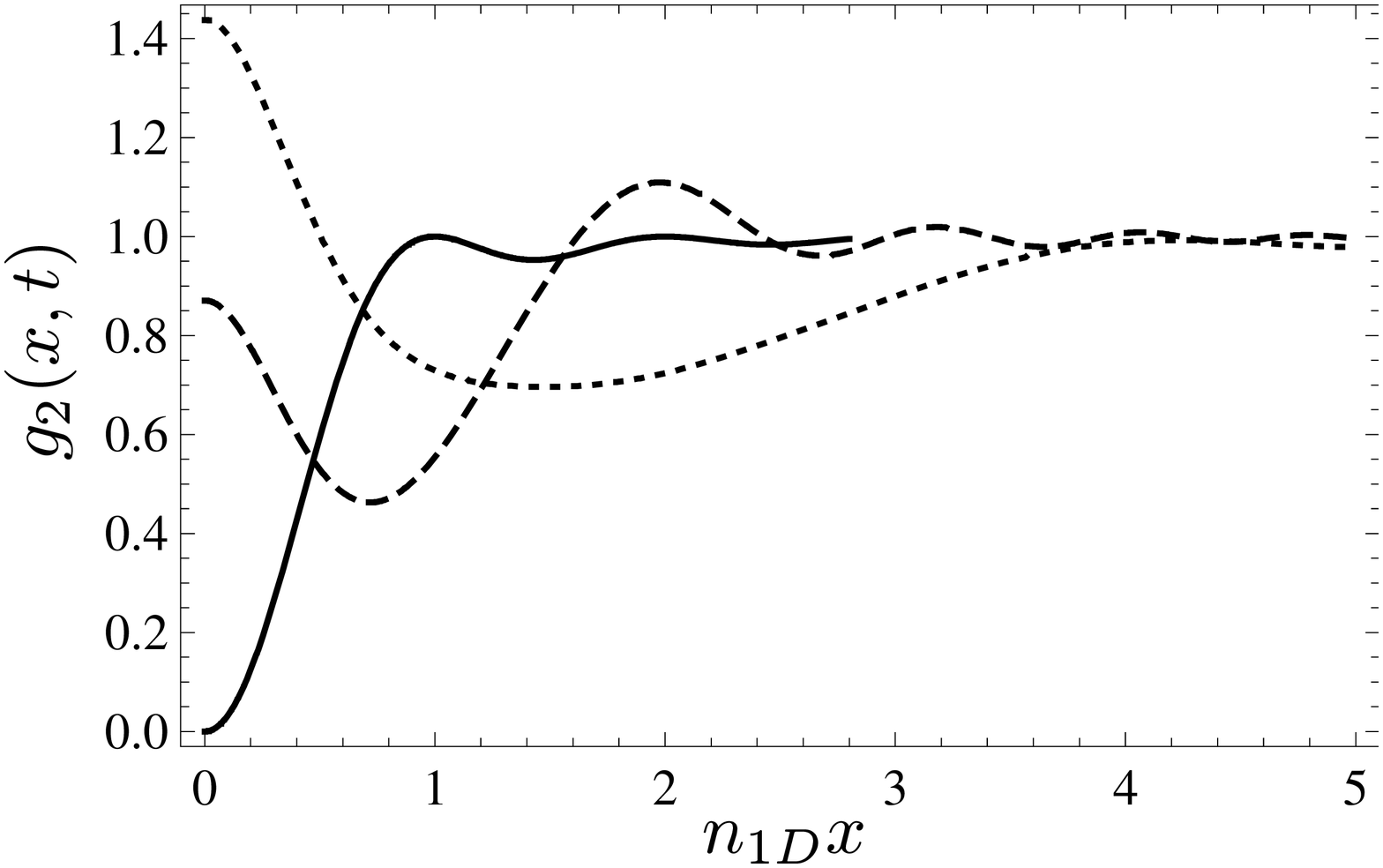}

\vspace*{3mm}

\caption{ \label{TG} Two-point density  correlation function
$g_2(x;t)$  of a zero-temperature strongly interacting 1D Bose gas
(Tonks-Girardeau limit) for different times $t$ after the release
of the gas from the trap. Different curves correspond to $t=0$
(solid), $t=0.25 m/(\hbar n_{1D}^2)$ (dashed),  and $t=m/(\hbar
n_{1D}^2)$ (dotted).
}
\end{figure}

\begin{figure}

\vspace*{3mm}

\includegraphics[width=8.5cm]{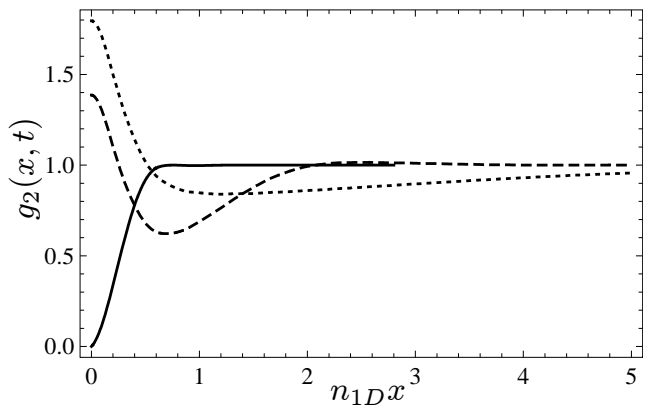}

\vspace*{3mm}

\caption{ \label{TGnzt} Two-point density  correlation function
$g_2(x;t)$   for the same parameters as in Fig.~\ref{TG} but for
a finite temperature, $k_BT=\mu\approx1.2\frac{\left (\pi \hbar
n_{1D}\right)^2}{2m}.$}
\end{figure}

In the Appendix  we derive a representation of
$\varrho(x_1;x_2;0)$ as a Fredholm-type determinant, which can be
easily evaluated numerically. Combining this representation with
Eq.~(\ref{eq:10}), we evaluate $g_2(x;t)$ after various expansion
times numerically. The results for zero temperature are shown in
Fig.~\ref{TG}, while the results for finite temperature $k_BT =
\mu\approx1.2\frac{\left (\pi \hbar n_{1D}\right)^2}{2m}$ are
shown in Fig.~\ref{TGnzt}. In spite of a considerable change in
the temperature, there is no qualitative change in the behavior
of $g_2(x;t).$ The qualitative behavior of $g_2(x;t)$ in
Figs.~\ref{TG} and~\ref{TGnzt}  is in agreement with
Eq.~(\ref{eq:11}) below, and $\lambda _C\sim n_{1D}^{-1}$ for the
Tonks-Girardeau gas.

\subsection{General remarks about 1D case}

\label{1Dgas}

Before concluding this section we would like to provide a
qualitative analysis of the evolution of the density correlation
function $g_2(x;t)$ as a function of the expansion time $t$.

\begin{figure}

\vspace*{3mm}

\includegraphics[width=8.5cm]{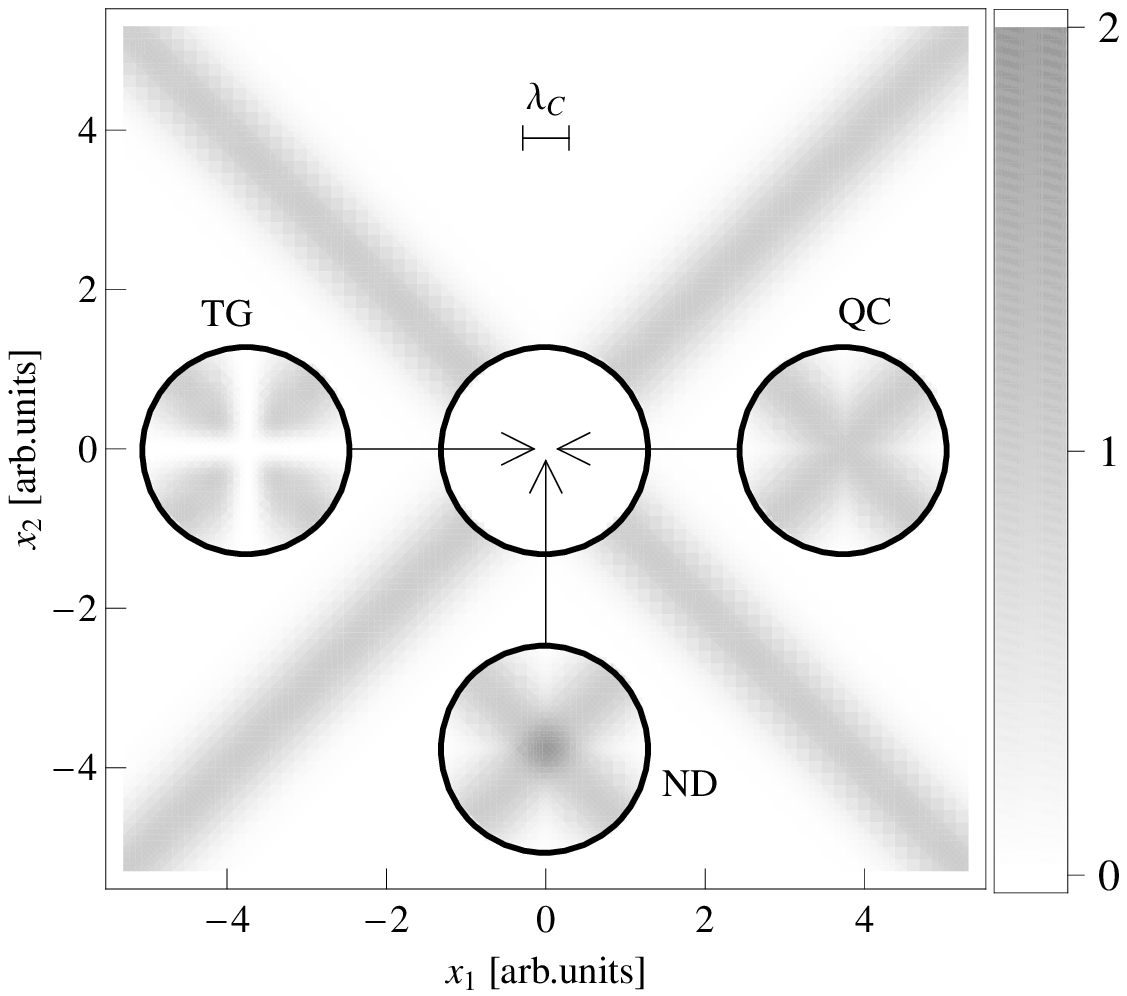}

\vspace*{3mm}

\begin{caption}
{\label{densityplot} Density plot of the two-particle density
matrix $\varrho (x_1;x_2;0)$ of a 1D Bose gas, see
Eqs.~(\ref{eq:4}) and (\ref{eq:7}) (the density bar represents the
$\varrho $ scale in units of $n_{1D}^2$). Initially (at $t=0$)
the bosonic system can be a Tonks-Girardeau gas (TG) or a weakly
interacting quasicondensate (QC) or a non-degenerate (ND) thermal
gas. The central ($x_1\approx x_2\approx 0$) part of the
two-particle density matrix in these cases is shown in three
respective insets. The bar shows the typical correlation scale
$\lambda _C$.}
\end{caption}
\end{figure}

The general structure of the two-particle density matrix $\varrho
(x_1;x_2;0)$ of a 1D Bose gas is shown schematically in
Fig.~\ref{densityplot}. Because of the Bose symmetry, $\varrho
(x_1;x_2;0)$ is represented in the $(x_1,x_2)$-plane by two
infinite perpendicular ``bands'' of a typical transverse size
$\lambda _C$ (correlation length). Asymptotically, as
$x_1\rightarrow \pm \infty$ and $x_2=\pm x_1$, $\varrho
\rightarrow n_{1D}^2$. There are several possible cases of atomic
correlations near the point $x_1=x_2=0$ in a trapped 1D gas. In
general,  at $t=0$, we have $\rho (0;0;0)= n_{1D}^2g_2(0;0)$. In
the case of the Tonks-Girardeau gas of impenetrable bosons
\cite{Gir,g1TG} $g_2(0;0)\equiv g_2^{TG}(0)=0$ (at zero
temperature $g_2^{TG}(x)=1-[\sin (\pi n_{1D}x)/(\pi n_{1D}x)]^2$
\cite{Gir}). Another possibility is a weakly interacting
degenerate gas (quasicondensate), where $g_2(0;0)\equiv
g^{QC}_2(0)\approx 1$ \cite{MoraCastin,g2first,asg,g2last,Deuar}.
Finally, the 1D Bose gas can be non-degenerate (thermal), in
which case $g_2(0;0) \equiv g_2^{ND}=2$. As the interparticle
distance grows, the density correlation function quite rapidly
approaches its asymptotic value $g_2(x \rightarrow \infty;0)=1$ at
the distances of the order of $\lambda_C.$
%

One can show that  time-dependent density correlation function
can be written  as
\begin{equation}
g_2(x;t)=1+\kappa(\lambda _C,x,t)+[g_2(0;0)-1]h(\lambda _C,x,t).
\label{eq:11}
\end{equation}
The first term (unity) stems from the band  of non-zero values of
$\varrho $ aligned along the line $x_1=x_2$ (see
Fig.~\ref{densityplot}). It represents the density correlation
function  of an ideal gas of distinguishable particles at
equilibrium. The second term, $\kappa(\lambda _C,x,t)$, reflects
the Bose-Einstein statistics of the atoms and appears due to the
second band along $x_2=-x_1$. Its maximum value, $\kappa(\lambda
_C,0,t)$, increases from 0 to 1 on a typical time scale $\sim
m\lambda _C^2/\hbar $. As $|x|$ grows, this term asymptotically
approaches 0 on a length scale given by $\lambda_C.$ The third
term describes washing-out of initial short-range (microscopic)
correlations. The maximum value of $h(\lambda _C,x,t)$ is reached
at $x=0$, it decreases from 1 to 0 on a time scale $\sim m\lambda
_C^2/\hbar $, and $h(\lambda _C,x,t)\approx 0$ if $|x|\gg \lambda
_C$. In the course of free evolution, the density correlation
properties of an expanding Bose gas become similar to that of an
ideal Bose gas at temperature $k_BT\sim \hbar ^2/(m\lambda _C^2)$.

\section{2D Bose gases below the Berezinskii-Kosterlitz-Thouless temperature}
\label{2Dsection}

Let us now discuss the properties of density ripples in expanding
2D clouds. Recently 2D condensates have been realized
experimentally in several
groups~\cite{gorlitz,Zoran_KT,Clade,2dconds,Kruger,JILA}. Reduced dimensionality
has dramatic effect on thermal fluctuations. In the case of 2D Bose
gases  there is no true long-range order for any finite temperature ~\cite{MWH}.
For uniform 2D Bose clouds at sufficiently low temperatures, the two-point
correlation function behaves at large distances
as~\cite{Kadanoff,Popov,Petrov2D_PRL,Petrov_review}
 \bea
\left \langle  \hat{\psi }^\dag ({\bf r},0) \hat{\psi }(0,0)
\right \rangle \approx n_{2D}
\left(\frac{\lambda_{2D}}{r}\right)^{\eta} \; \mbox{for} \;
r\gg\lambda_{2D}. \label{2DCorr}
 \eea
For weakly interacting 2D Bose gas at small temperatures, one can
evaluate parameters of Eq.~(\ref{2DCorr})  from microscopic
theory.  The dimensionless parameter characterizing weakness of
interactions is written as~\cite{Petrov2D_PRL,Petrov_review}
 \bea
 \tilde g=a_s\sqrt{\frac{8\pi m \omega_{\perp}}{\hbar}} \ll1. \label{gtilde}
 \eea
The exponent $\eta$ in Eq.~(\ref{2DCorr})
equals~\cite{Petrov2D_PRL,Petrov_review}
 \bea
 \eta=\frac{T}{T_d} \ll 1 \; \; \mbox{for} \;k_B T\ll k_B T_d=\frac{2\pi \hbar^2
 n_{2D}}{m}, \label{2Dweakexp}
 \eea
and $\lambda_{2D}$ equals the de Broglie wavelength of thermal
phonons $\hbar c/(k_BT)$ at $k_B T\ll \mu, $ and the
two-dimensional healing length $\xi_{2D}=\hbar/\sqrt{m\mu}$ at
high temperatures $k_B T\gg \mu.$

Equation (\ref{2DCorr}) remains valid for $\eta$ smaller than
 \bea
 \eta_c=1/4,
 \eea
at which point the BKT~\cite{Berezinskii,KT,GNT} transition takes
place due to proliferation of vortices, and correlation functions
start to decay exponentially with distance.

Such a  transition for ultracold 2D Bose gases has been observed
recently~\cite{Zoran_KT,Clade,JILA}, and its microscopic origin
has been elucidated. Experiments of Ref.~\cite{Zoran_KT} studied
interference of two independent 2D Bose clouds, which requires
imaging along the ``in-plane'' direction and inevitably leads to
averaging over inhomogeneous densities. Study of the spectrum of
density ripples in expanding clouds with imaging in transverse
direction (as done in Ref.~\cite{Clade}) avoids this problem
altogether, and can provide access to properties of correlations
at fixed density. The interplay between the BKT transition and
the effects of the external confinement is a rather complicated
question
even for weakly interacting Bose
gas~\cite{Kruger,Hadzi_NJP,Holzmann,2DRMP}, and  we will only
discuss the uniform case here.

Even for weak interactions, one cannot use quasicondensate theory
to analytically describe correlations as functions of microscopic
parameters in the vicinity of the BKT transition or to predict
the transition temperature, and has to resort to fully numerical
methods~\cite{Prokofiev}. Nevertheless, the factorization property
[Eq.~(\ref{factor})] remains valid for large-distance behavior of
correlation functions for all $\eta$ below the critical value
$1/4,$ since large-distance fluctuations of the phase are still
described by the Gaussian theory.  Using that together with
Eq.~(\ref{2DCorr}), we will now obtain the prediction for the
spectrum of density ripples which is valid as long as only points
with relative distances much larger than $\lambda_{2D}$
contribute significantly to the integral in Eq.~(\ref{rhoqeq2}).
We will show below, that this regime is realized if
 \bea
 \sqrt{\frac{\hbar t}{m}} \gg \lambda_{2D}. \label{2Dsc}
 \eea
We introduce a dimensionless variable
 \bea
 y =\frac{\hbar q^2 t }{m},
 \eea
and use expression Eq.~(\ref{2DCorr}) for all ${\bf r}.$ Using
symmetries of the resulting integral, the expression for $\left
\langle |\rho(q)|^2\right \rangle$ is written as
 \bea
 \left \langle |\rho(q)|^2\right \rangle\approx n_{2D}^2
 \lambda_{2D}^2 \left(\frac{\hbar t }{m \lambda^2_{2D}}\right)^{1-\eta}
 F(\eta,y), \label{scalingans}
 \eea
where $F(\eta,y)$ is a dimensionless function defined by
~\cite{fnote}
 \bea
 F(\eta,y)= \frac{4}{y^{1+\eta}}\int_{0}^{\infty} dr_x
 \cos{r_x}\int_{0}^{\infty}dr_y\nonumber \\
 \left\{\left[\frac{\sqrt{(r_x+y)^2+r_y^2}
 \sqrt{(r_x-y)^2+r_y^2}}{r_x^2+r_y^2}\right]^\eta-1\right\}.
 \label{Fdef}
 \eea

\begin{figure}

\vspace*{3mm}
\includegraphics[width=8.5cm]{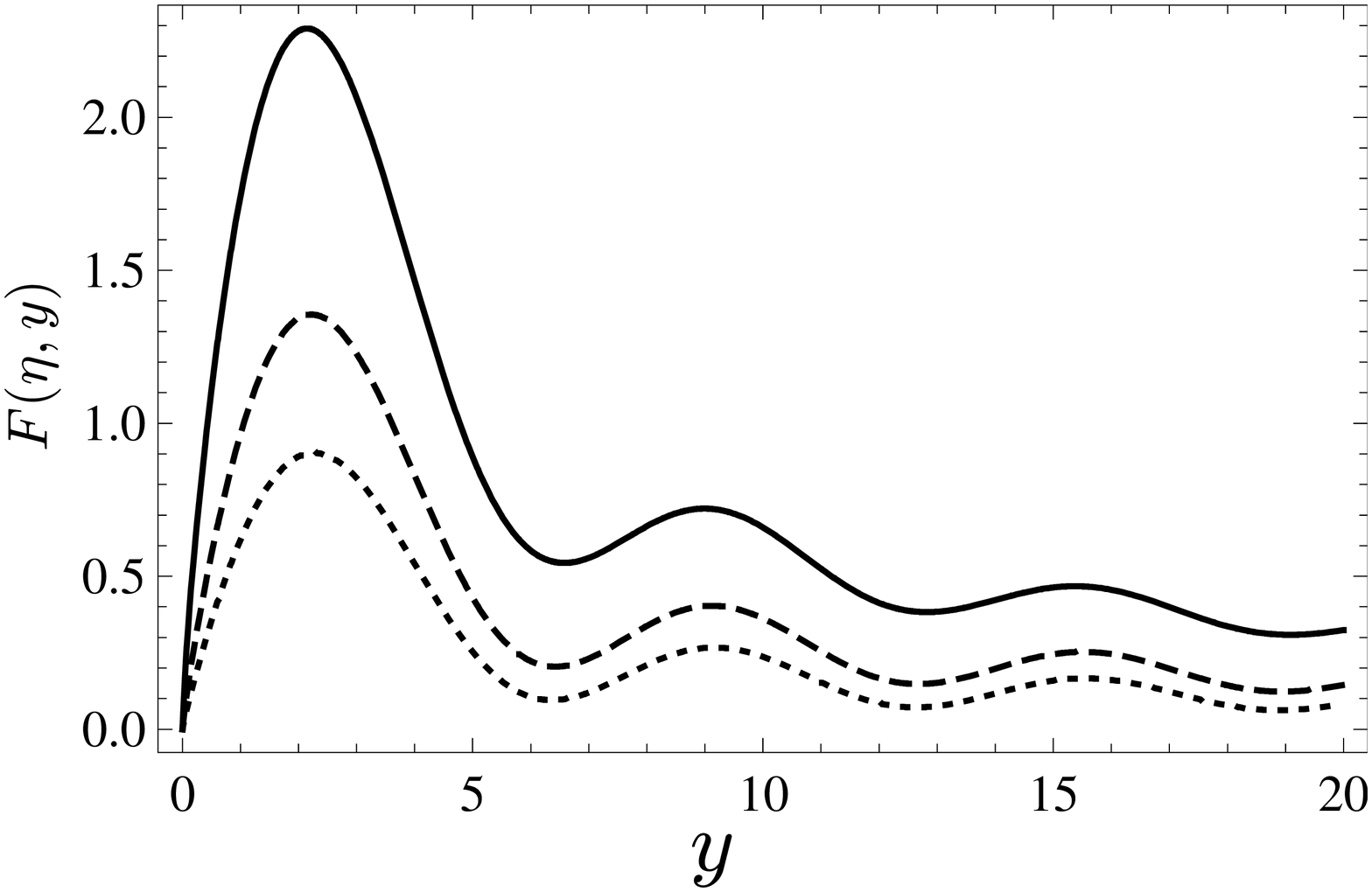}
\vspace*{3mm}

\caption{\label{Fetaplot} Dependence of universal functions
$F(\eta,y)$ on $y=\frac{\hbar q^2 t}{m}$ plotted for three
different values of correlation exponents $\eta.$ Under condition
(\ref{2Dsc}) functions $F(\eta,y)$ determine the self-similar
shape of the spectrum of density ripples according to
Eq.~(\ref{scalingans}). Curves from top to bottom correspond to
$\eta= 0.25$ (solid, the Berezinskii-Kosterlitz-Thouless point),
$\eta= 0.15$ (dashed), and $\eta= 0.10$ (dotted).}
\end{figure}

 We find
that the spectrum of density ripples remains self-similar in the
course of expansion  and the shape of the spectrum is a function
of $\eta$ only. Plots of $F(\eta,y)$ for three different values
of $\eta$ are shown in Fig.~\ref{Fetaplot}, and have a similar
structure. Positions of maxima and minima are very well described
by Eq.~(\ref{extrrule}), where the upper (lower) sign corresponds
to the $n$th maximum (minimum). In Eq.~(\ref{rhoqeq2}) typical
distances which contribute to $\left \langle |\rho(q)|^2\right
\rangle$ near its maximum at $y\approx \pi$ can be estimated as
$\sim \sqrt{\hbar t/m},$ which leads to condition (\ref{2Dsc}).
Note however, that self-similarity starts breaking down for
sufficiently large $y$ even when condition (\ref{2Dsc}) is
satisfied.

Scaling of the magnitude of  $\left \langle |\rho(q)|^2\right
\rangle$ with time in the self-similar regime can be used to
extract $\eta.$ For example, the integral of $\left \langle
|\rho(q)|^2\right \rangle$ from zero to its  first minimum  scales
with time  as
 \bea
  \int_{0}^{\sqrt{\frac{2\pi m}{\hbar t}}} dq  \left\langle |\rho(q)|^2\right
  \rangle\propto t^{1/2-\eta}, \label{scaling}
 \eea
and the exponent changes considerably as $\eta$ changes from $0$
to the critical value $1/4.$

For small $\eta,$ one can derive an expansion of $F(\eta,y)$ as
 \bea
 F(\eta,y)=\frac{4}{y^{1+\eta}}\left[\eta f_1(y) + \eta^2 f_2(y)+\eta^3 f_3(y)+...\right],
 \eea
where $f_1(y)$ can be evaluated analytically as
 \bea
 f_1(y)=2\pi \sin^2{\frac{y}{2}}.  \label{Petrovmagic}
 \eea
The term $f_2(y)$ leads to a finite value of $F(\eta,y)$ at the
first minimum. By including effects of $f_2(y)$ and $f_3(y),$ one
can derive
 \bea
 \frac{F(\eta,2\pi)}{F(\eta,\pi)}\approx\frac1{2^\eta}\left( 1.19\eta+0.38\eta^2\right) \; \mbox{for} \;
 \eta \ll 1,
 \eea
which coincides with the direct numerical evaluation up to
$2.5\%$ for $\eta=0.25.$

For weakly interacting uniform 2D Bose gases at low temperatures,
one can also obtain predictions which are not limited by
Eq.~(\ref{2Dsc}).
Under condition
 \bea
 n_{2D} \xi_{2D}^2\gg1
 \eea
an extension of Bogoliubov theory to 2D quasicondensates describes
correlations at all distances~\cite{MoraCastin}. Such a  theory is
valid up to temperatures of the order
 \bea
 \frac{k_B T}{\mu} \log{\frac{k_B T}{\mu}} \sim n_{2D} \xi_{2D}^2
 \gg1, \label{2Dweak}
 \eea
 and predicts the exponent~(\ref{2Dweakexp}).
 The correlation function is written as
 \bea g_1({\bf r})=\frac{\left \langle  \hat{\psi
}^\dag ({\bf r},0) \hat{\psi }(0,0)\right
\rangle}{n_{2D}}=\exp{\left[-\frac{2\pi \mu}{ k_B T_d
}f_{2D}\left(\frac{r}{\xi_{2D}}\right)\right]}, \nonumber
 \eea
where the dimensionless function $f_{2D}(s)$ is defined by
 \bea
 f_{2D}(s)=\int_{0}^{\infty}\frac{k dk}{2\pi} \left[1-J_0({k s})\right] \left\{\right[u_k^2+v_k^2\left] n_k + v_k^2
 \right\}.\nonumber
 \eea
Here, $J_0(x)$ is the Bessel function, and $u_k,v_k$ and $n_k$ are
defined by Eqs.~(\ref{17})-(\ref{19}).

\begin{figure}

\vspace*{3mm}
\includegraphics[width=8.5cm]{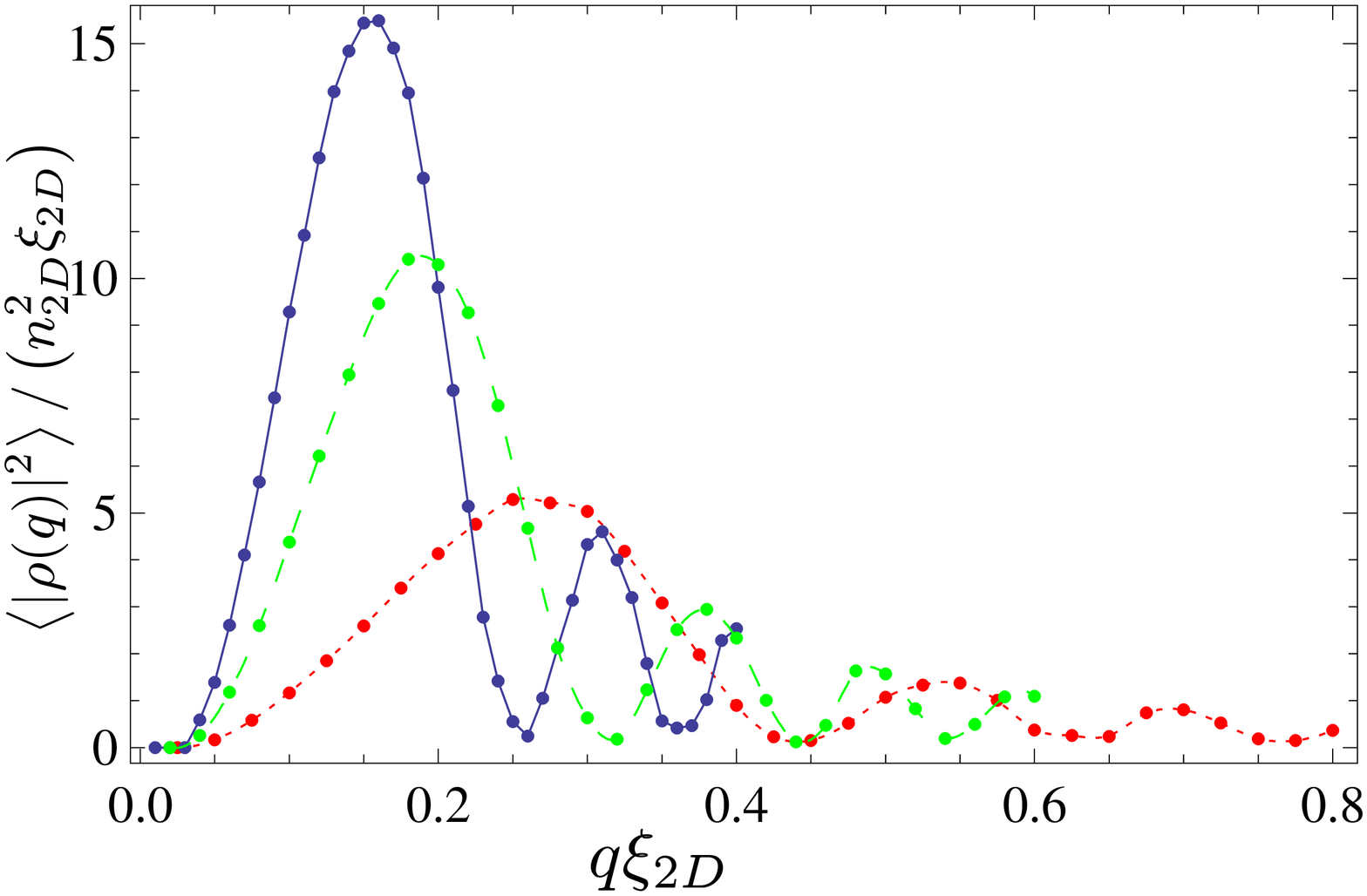}
\vspace*{3mm}

\caption{\label{2Dweakplot}(Color online) Normalized spectrum of
density ripples $\left \langle|\rho(q)|^2\right \rangle/(n_{2D}^2
\xi_{2D})$ for weakly interacting 2D quasicondensate of $^{87}$Rb
atoms with density $n_{2D}\approx 84\,\mbox{$\mu$}\mbox{m}^{-2},$
transverse confinement frequency $\omega_\perp=2\pi \times
3\,\mbox{kHz},$ healing length $\xi_h\approx
0.3\,\mbox{$\mu$}\mbox{m},$ and dimensionless interaction
parameter $\tilde g\approx 0.13.$ Temperature is taken to be
$T=60\,\mbox{nK}\, (k_BT\approx \mu),$ which corresponds to
correlation exponent $\eta=0.020.$ Various curves correspond to
expansion times $t=12\,\mbox{ms}$ (blue, solid), $ t=8\,\mbox{ms}$
(green, dashed), and $t=4\,\mbox{ms}$ (red, dotted). The lines
are guides to the eyes. }
\end{figure}

We now consider a case of $^{87}$Rb atoms with transverse
confinement frequency $\omega_\perp=2\pi \times 3\,\mbox{kHz},$
and density $n_{2D}\approx 84\,\mbox{$\mu$}\mbox{m}^{-2}.$ This
yields the dimensionless interaction parameter $\tilde g\approx
0.13,$ healing length $\xi_h\approx 0.3\,\mbox{$\mu$}\mbox{m},$
and $n_{2D} \xi_{2D}^2\approx 7.5.$ We perform a numerical
integration of Eq.~(\ref{rhoqeq2}) for temperature
$T=60\,\mbox{nK}\,$(which corresponds to correlation exponent
$0.02$) and various expansion times, and the results are shown in
Fig.~\ref{2Dweakplot}.

Qualitatively, they look similar to the self-similar regime for
all times, and one again obtains an oscillating spectrum of
density ripples with maxima and minima very well described by
Eq.~(\ref{extrrule}). In the weakly interacting regime the ratio
of the first maximum to the first minimum for $\left
\langle|\rho(q)|^2\right \rangle$ is much larger than one,
similar to the weakly interacting 1D Bose gas.

\section{Conclusions}
\label{conclusions}

To conclude, we calculated the evolution of the two-point density
correlation function of an ultracold atomic Bose gas released
from a tight transverse confinement.

For 1D  gases in the weakly interacting regime, in a wide
range of parameters given by Eq.~(\ref{analcond}), we analytically
calculated the spectrum of density ripples $\left \langle
|\rho(q)|^2\right \rangle.$ Our results are summarized
in  Eq.~(\ref{rhoqfinal}) and
Figs.~\ref{TP} and \ref{TF}.
Our analytical theory is also applicable in the quasi-1D regime
when $k_BT$ and $ \mu$ are of the order of transverse confinement
frequency $\hbar \omega_\perp.$ For  expansion times smaller than
$ 6.5 m \lambda_T^2/\hbar,$  we find that the spectrum of density
ripples can have several maxima and minima, and their positions
can be estimated using Eq.~(\ref{extrrule}). While positions of
maxima and minima are essentially independent of  the
temperature, their amplitude  exhibits strong temperature
dependence. For 1D quasicondensates, the density profile in
external harmonic confinement depends weakly  on the temperature,
when the latter is of the order of the chemical
potential~\cite{MoraCastin}. The density profile follows the
inverted parabola shape~\cite{Pollet}, thus the bimodal density
fitting cannot be used to  measure temperatures reliably. We
propose that our analytical result Eq.~(\ref{rhoqfinal}) can be
used for thermometry of one-dimensional systems. Experimental
investigation of this question is currently under way, and will
be presented in a separate presentation~\cite{Manz}.

For one-dimensional systems, we also discussed evolution of the
density correlation function in real space, $g_2(x;t)$. For long
expansion times we find that the correlation function $g_2(x;t)$
reaches the value of $2$ at short distances and approaches the
value $1$  for distances larger than the correlation length, see,
e.g., Fig.~\ref{TG} for Tonks-Girardeau regime.

For 2D Bose gases with temperatures below the
Berezinskii-Kosterlitz-Thouless transition and sufficiently long
expansion time, we showed that the spectrum of the density ripples
evolves in a self-similar way. Our result for this case is given
in Eq.~(\ref{scalingans}) and Fig.~\ref{Fetaplot}, with positions
of maxima and minima determined by Eq.~(\ref{extrrule}). The
scaling of the overall magnitude can be used to extract the
correlation exponent $\eta,$ e.g., using Eq.~(\ref{scaling}).



For more complicated situations, e.g., multicomponent gases,
relation (\ref{rhoqeq}) and its cross-correlation generalizations
can be used as a convenient experimental tool to characterize
complex many-body states and their correlations. In addition, it
can be used as an experimental tool to investigate
non-equilibrium phenomena in low-dimensional gases.


A.I. was supported by DOE Grant No.~DE-FG02-08ER46482. I.E.M.
acknowledges support through the Lise Meitner program by the FWF,
and the INTAS. D.S.P. was supported by  ANR Grant
No.~08-BLAN-0165-01 and by the Russian Foundation for Fundamental
Research. V.G. was supported by Swiss National Science Foundation.
S.M. acknowledges support from the FWF doctoral program CoQuS.
T.S. acknowledges support by
the FWF program P21080.
 E.D. was supported by the NSF Grant No.~DMR-0705472, DARPA, MURI, and Harvard-MIT CUA.
J.S. was supported by the EC, and the FWF. We acknowledge useful
discussions with A.~Aspect, P.~Clad\'{e}, E.~Cornell, J.~Dalibard,
N.~J.~van~Druten, M.~Greiner, Z.~Hadzibabic, K.~Kheruntsyan, and
W.~Phillips.

\appendix
\section{Two-particle density matrix of a strongly interacting 1D  Bose gas}
\label{AppA}

In this appendix we will describe a Fredholm-type determinant
representation for $\varrho(x_1;x_2;0)$ for $0<x_1<x_2,$ which can
be easily evaluated numerically. Due to Eqs.~(\ref{eq:8})
and~(\ref{eq:9}) this defines $\varrho(x_1;x_2;0)$ for any values
of $x_1$ and $x_2.$ Representations  similar to the one developed
here can be obtained for any multi-point correlation function of
bosonic fields in the strongly interacting limit.

Mathematically, fermionization can be written as
 \bea
 \hat \psi^{\dag} (x)= \exp{\left[i\pi \int^{x-0}_{-\infty}dy\, \hat\psi_f^{\dag}(y)\hat\psi_f(y)\right]}\hat\psi_f^{\dag}(x), \label{ferm1}\\
  \hat \psi(x)= \exp{\left[-i\pi \int^{x-0}_{-\infty}dy\, \hat\psi_f^{\dag}(y)\hat\psi_f(y)\right]}\hat\psi_f(x)\label{ferm2},
 \eea
where we introduced fermionic creation and annihilation operators
$\hat \psi_f^{\dag}(x)$ and $\hat  \psi_f(x), $ which have
standard anti-commutation relations
 \bea
  \left\{\hat\psi_f^{\dag}(x),\hat\psi_f(y) \right\}=\delta(x-y), \label{comm1} \\
  \left\{\hat\psi_f^{\dag}(x),\hat\psi^{\dag}_f(y) \right\}=  \left\{\hat\psi_f(x),\hat\psi_f(y)
  \right\}=0.\label{comm2}
 \eea
For zero temperature, ground state for fermions corresponds to a
filled Fermi sea, whereas at finite temperature one should use a
thermal density matrix for non-interacting fermions.

For convenience, we will introduce a fictitious underlying
lattice of spacing $a \ll x_1, x_2,$ such that
 \bea
 \frac{x_1}{2}\frac1{a}=m_1 \gg1,\\
 \frac{x_2}{2}\frac1{a}=m_2>m_1 \gg1,
 \eea
where $m_1$ and $m_2$ are large positive integer numbers. At the
end of the calculation, we will take the limit $a\rightarrow 0$
such that $m_1 a\rightarrow x_1/2, m_2 a\rightarrow x_2/2.$ On a
lattice, fermionization rules [Eqs.~(\ref{ferm1}) and
(\ref{ferm2})] and commutation relations (\ref{comm1}) and
(\ref{comm2}) are written as
 \bea
\hat \psi^{\dag} (i)= \prod_{k<i}\left[1-2\hat\psi_f^{\dag}(k)\hat\psi_f(k)\right]\hat\psi_f^{\dag}(i), \label{ferm3}\\
\hat \psi (i)= \prod_{k<i}\left[1-2\hat\psi_f^{\dag}(k)\hat\psi_f(k)\right]\hat\psi_f(i), \label{ferm4}\\
  \left\{\hat\psi_f^{\dag}(i),\hat\psi_f(k) \right\}=\delta_{ik}, \label{comm3} \\
  \left\{\hat\psi_f^{\dag}(i),\hat\psi^{\dag}_f(k) \right\}=  \left\{\hat\psi_f(i),\hat\psi_f(k)\right\}=0.\label{comm4}
 \eea
Using these relations, $\varrho(x_1;x_2;0)$ can be written as
 \bea
 \varrho(x_1;x_2;0)=\left\langle \hat\psi_f^{\dag}(m_1)\hat\psi_f^{\dag}(-m_1)  \right. \nonumber \\
\times\left. \prod_{k\in
S}\left[1-2\hat\psi_f^{\dag}(k)\hat\psi_f(k)\right]
\hat\psi_f(m_2)\hat\psi_f(-m_2)
 \right\rangle,
 \eea
where subset $S$ equals
 \bea
 S=\left[-m_2+1,-m_1-1\right] \bigcup \left[m_1+1,m_2-1\right].
 \eea
Expanding the parentheses, we obtain
 \bea
\varrho(x_1;x_2;0)=\left\langle\sum_{n=0}^{\infty}(-2)^n \hat\psi_f^{\dag}(m_1)\hat\psi_f^{\dag}(-m_1)  \right. \nonumber \\
\times \sum_{j_1<...<j_n, j_k\in S}
\hat\psi_f^{\dag}(j_1)...\hat\psi_f^{\dag}(j_n)\hat\psi_f(j_n)...\hat\psi_f(j_1)\nonumber
\\
\times\left. \hat\psi_f(m_2)\hat\psi_f(-m_2)
 \right\rangle.\label{sumrepr}
 \eea
For each $n$ and set of $j_1,...,j_n,$ expectation value of $n+2$
creation and $n+2$ annihilation operators can be written using
Wick's theorem~\cite{AGD} as a determinant of $(n+2)\times(n+2)$
matrix~\cite{Burovskii, Zvonarev_string}
 \bea
 M^{(n+2)}_{i,j}= a G(s_i,t_j),\label{Mij}
 \eea
where
 \bea
 s_1=-m_1 a, \; s_2=m_1 a, \;s_{i>2}=j_{i-2} a, \\
 t_1=m_2 a, \; t_2=-m_2 a, \;t_{i>2}=j_{i-2} a,
 \eea
and $G(x,y)=G(x-y)$ is a Green's function of a free Fermi gas,
which e.g. for zero temperature equals
 \bea
 G(x)=\int_{-k_f}^{k_f}\exp{\left[i k
 x\right]}\frac{dk}{2\pi}=\frac{\sin{\pi n_{1D}x}}{\pi
x}.
 \eea
Since the structure of the matrix $M^{(n+2)}_{i,j}$ does not
depend on $n,$ summation over different $n$ and  sets
$j_1,...,j_n$  can be now represented as a single Fredholm-type
determinant~\cite{Zvonarev_string, Smirnov_book}
 \bea
 \varrho(x_1;x_2;0)=\frac{\mbox{Det}\;\left[A_{ij}-2 a B_{ij}\right]}{4a^2},
 \eea
where matrices $A_{ij}$ and $B_{ij}$ of size $2(m_2-m_1)\times
2(m_2-m_1)$ are defined by
 \bea
 A_{ij} = \,\mbox{Diag} \{0,0,1,...,1\},\\
 B_{ij} = G(\tilde s_i, \tilde t_j),
 \eea
and
 \bea
 \tilde s_1=-m_1 a, \; \tilde s_2=m_1 a,\\
 \tilde s_i=\left(-m_2+i-2\right) a \; \mbox{for} \; 3\leq i<2+m_2-m_1, \\
 \tilde s_i=\left(2m_1-m_2+i-1\right) a \; \mbox{for}\nonumber \\ \; 2+m_2-m_1\leq i\leq 2(m_2-m_1), \\
 \tilde t_1=m_2 a, \; \tilde t _2=-m_2 a, \;\tilde t_{i>2}=\tilde s_{i}.
 \eea
Expansion of the determinant of $A_{ij}-2 a B_{ij}$ using the
rule for the determinant of the sum of two matrices (see, e.g.,
p.~221 of Ref.~\cite{KBI}) generates the expansion of
Eq.~(\ref{sumrepr}), similar to a usual Fredholm
determinant~\cite{Smirnov_book}. Indeed, only diagonal minors not
including lines $1$ and $2$ can be chosen from the matrix
$A_{ij}.$ Complimentary minor of size $(n+2)\times(n+2)$ from the
matrix $B_{ij}$ is proportional to matrix $M^{(n+2)}$ in
Eq.~(\ref{Mij}), and the summation over possible different sets of
$j_1,...,j_n$ is equivalent to a  summation over different
partitions of matrix $A_{ij}$ into diagonal minors.

Since determinants are easy to evaluate numerically, one can now
take the limit $a\rightarrow 0$ numerically and evaluate
$\varrho(x_1;x_2;0)$ with any precision.


\begin{thebibliography}{99}
\bibitem{positivitynote} One should note that, strictly speaking, $\left \langle |\rho(q)|^2\right
\rangle$ does not have to be positive, since it is not an
expectation value of a positive operator. According to our
definition [Eq.~(\ref{rhovsg2})] square of the operator $\int dx
\exp{\left(i q x \right)}\hat{\psi }^\dag (x,t)\hat{\psi}(x,t)$
differs from $L \left \langle |\rho(q)|^2\right \rangle$ by a
term which arises due to normal ordering present in the
definition of $g_2(x;t).$
\bibitem{minus1note} In the right-hand side of Eq.~(\ref{rhovsg2}) the term $-1$ in
the parentheses was introduced to compensate for the large
distance behavior of $g_2(x;t),$ and does not affect  the
spectrum at $q\neq0.$
\bibitem{fnote} Constant $-1$ in Eq.~(\ref{Fdef}) is included for the
convenience to make the integrand vanish at $r_y\rightarrow
\infty,$  and its contribution disappears after integration over
$r_x.$
\end{thebibliography}
\end{document}